# Native point defects and low *p*-doping efficiency in Mg$_2$(Si,Sn) solid solutions: A hybrid-density functional study

**Byungki Ryu,[1],* Eun-Ae Choi,[2] Sungjin Park,[1] Jaywan Chung,[1] Johannes de Boor,[3] Pawel Ziolkowski,[3] Eckhard Müller,[3,4] and SuDong Park[1]**

[1] *Energy Conversion Research Center, Electrical Materials Research Division, Korea Electrotechnology Research Institute (KERI), Changwon 51543, Republic of Korea*

[2] *Computational Materials Department, Materials Processing Innovation Research Division, Korea Institute of Materials Science (KIMS), Changwon 51508, Republic of Korea*

[3] *German Aerospace Center (DLR) – Institute of Materials Research, Cologne 51147, Germany*

[4] *Institute of Inorganic and Analytical Chemistry, Justus Liebig University, Giessen, 35392 Giessen, Germany*

# ABSTRACT

We perform hybrid-density functional calculations to investigate the charged defect formation energy of native point defects in $Mg_2Si$, $Mg_2Sn$, and their solid solutions. The band gap correction by hybrid-density functional is found to be critical to determine the charged defect density in these materials. For $Mg_2Si$, Mg interstitials are dominant and provide unintentional *n*-type conductivity. Additionally, as the Mg vacancies can dominate in Mg-poor $Mg_2Sn$, *p*-type conductivity is possible for $Mg_2Sn$. However, the existence of low formation energy defects such as $Mg_{Sn}^{1+}$ and $I_{Mg}^{2+}$ in $Mg_2Sn$ and their diffusion can cause severe charge compensation of hole carriers resulting in low *p*-type doping efficiency and thermal degradation. Our results indicate that, in addition to the extrinsic doping strategy, alloying of $Mg_2Si$ with $Mg_2Sn$ under Mg-poor conditions would be necessary to enhance the *p*-type conductivity with less charge compensation.

**1. Introduction**

Mg$_2$Si is a potential semiconductor for thermoelectric and optoelectronic devices owing to its high power factor and high optical absorption coefficient, respectively [1-3]. By alloying with Mg$_2$Sn, the performance of the materials can be tuned to achieve high thermoelectric performance via conduction band convergence [4] or control the optical-absorption spectra by the band gap ($E_g$) engineering [3]. For such applications, the symmetric doping nature is required to form a *p-n* dual-leg in thermoelectric devices or a *p-n* junction in electrical devices [1,3].

There have been a lot of experimental investigations on the electrical properties of Mg$_2$Si-based alloys. Mg$_2$Si has been reported to have an unintentional *n*-type conductivity [5-8]. By alloying Mg$_2$Si with Mg$_2$Sn, with the narrower $E_g$ and the increased electrical conductivity, conduction band convergence, and enhanced phonon scattering by point disorders, the solid solutions can achieve higher *n*-type thermoelectric performance [4,9]. Although Mg$_2$Sn can show *p*-type conductivity [10,11], both Mg$_2$Si and Mg$_2$(Si,Sn) solid solutions are difficult to be doped as *p*-type even under Mg-poor conditions [2,12]. The hole carrier density in doped Mg$_2$(Si,Sn) is lower than the optimal doping concentration for an optimal power factor; thus, there is strong desires to reach high *p*-type doping efficiency for high thermoelectric performance.

Previous first-principles calculations reveal that native point defects in Mg$_2$Si and Mg$_2$Sn may play a critical role in determining their electrical properties. The Mg interstitials are a dominant defect and are responsible for unintentional *n*-type conductivity [13-15]. The origin of the *p*-type doping difficulty in Mg$_2$(Si,Sn) has been explained by the interplay between acceptors and intrinsic defects [2,15-18]. However, while the $E_g$ underestimation in density functional theory (DFT) is challenging in defect physics [19], many of these recent defect computations in thermoelectric materials are merely based on this scheme using conventional local-density approximation or generalized-gradient approximation (GGA). For example, as the Mg$_2$Sn is negative in DFT calculations [20], the detailed quantitative role of intrinsic defects in Mg$_2$Sn and Mg$_2$(Si,Sn) solutions are not clearly understood. Furthermore, there have been many attempts outside the thermoelectric community, to calculate the defect properties using *E*$_g$ *correction methods* such as hybrid-DFT and quasi-particle GW calculations to reveal the device instability or doping asymmetricity for Si, ZnO, and HfO$_2$ [21-23]. Unfortunately, in contrast to other materials, there are few hybrid-DFT studies of native defects in Mg$_2$Si, Mg$_2$Sn, and their solid solutions. Moreover, to the best of our knowledge, there has been no hybrid-functional study on native defects in Mg$_2$Sn.

In this study, we investigated the native point defects in Mg$_2$Si, Mg$_2$Sn, and their solid solutions using hybrid-DFT within the Heyd-Scuseria-Ernzerhof (HSE) non-local exchange-correlation functional. The *E*$_g$ correction is found to be critical to determine the stability of positively charged defects especially for *p*-type conditions. Mg$_2$Si is unintentional *n*-type due to the Mg interstitial (I$_{Mg}^{2+}$). Mg$_2$Sn can be either *p*- or *n*-type depending on the Mg chemical potential. Although the Mg vacancies (V$_{Mg}^{2-}$) are dominant defects as a shallow acceptor for Mg$_2$Sn, many free hole carriers can be compensated by the Sn-substitutional defect at the Mg site (Sn$_{Mg}^{1+}$), leading to low *p*-doping efficiency. When Mg$_2$Si is alloyed with Mg$_2$Sn, *p*-type conductivity can be achieved in Mg-poor conditions. However, the overall free hole densities are not high enough to reach the optimal carrier concentration ($10^{20}$ cm$^{-3}$) to maximize the power factor. Our result suggests that, in addition to the material alloying between Mg$_2$Si and Mg$_2$Sn, external impurity point defect doping with very low formation energy might be necessary to achieve high *p*-type doping concentration with less charge compensation.

## 2. Computational Method

Our first principles calculations of native point defects in $Mg_2Si$ and $Mg_2Sn$ are based on hybrid-DFT within the HSE hybrid exchange-correlation functional (HSE06) [24], implemented in the VASP code [25,26]. We used the exact-exchange mixing parameter of 25% and the screening parameter of 0.208 Å$^{-1}$ with the GGA parameterized by Perdew, Burke, and Ernzerhof [27] and the projector-augmented wave (PAW) pseudopotentials [26]. Experimental lattice parameters were used for $Mg_2Si$ (6.35 Å) and $Mg_2Sn$ (6.75 Å) [8]. Note that, for HSE06 calculations, the optimized lattice parameters are calculated to be 6.33 Å and 6.765 Å for $Mg_2Si$ and $Mg_2Sn$, respectively. In conventional DFT the band gap problem is severe: for $Mg_2Sn$ the DFT $E_g$ is negative (-0.341 eV) whereas the experimental gap is about 0.36 eV [8,20]. As the charge state of the defects are sensitive to the Fermi level ($E_{Fermi}$), the positions of band edge states are critical for the defect stability. In this study, by adopting hybrid-DFT, reliable $E_g$s were obtained as 0.571 and 0.145 eV for $Mg_2Si$ and $Mg_2Sn$ respectively. We modeled the native point defects in $Mg_2Si$ and $Mg_2Sn$ using a 96-atom cubic supercell. We considered six defect configurations as native point defects: Mg interstitial ($I_{Mg}$), Mg vacancy ($V_{Mg}$), Mg substitutional at X site ($Mg_X$), X interstitial ($I_X$), X vacancy ($V_X$), X substitutional at Mg site ($X_{Mg}$) in $Mg_2X$ where X is Si or Sn. We used a plane-wave basis set with an energy cutoff of 296 eV and a 2 × 2 × 2 Γ-centered $k$-point mesh for integrations over the Brillouin zone. Atomic positions were fully relaxed using HSE06 level.

**Formation Energy.** The charged defect formation energies ($E_{FORM}$) of a defect in $Mg_2X$ with charge state $q$ ($d^q$), where q is 2+, 1, 0, 1-, 2-, were computed using the total energies of defective supercell and pristine supercell ($E_{tot}[d^q]$ and $E_O$), the atomic chemical potential ($\mu_i$), and the $E_{Fermi}$ given

as

$$E_{\text{FORM}}(\mu_i, E_{Fermi}|d^q, \text{Mg}_2\text{X}) = E_{\text{tot}}[d^q] - E_0 - \Sigma_i(\mu_i \delta n_i) + q(E_{\text{Fermi}} + E_{\text{CBM}})$$

where the subscript $i$ indicates the atomic element consisting pristine $Mg_2X$, $\delta n_i$ is the change of number of $i$ element atoms in the defective supercell compared to the pristine supercell, and $E_{\text{CBM}}$ is the energy of the conduction band minimum (CBM) state [28,29]. Note that the $E_{\text{Fermi}}$ can range from near the valence band maximum (VBM) to near the CBM. Considering the nearly preserved CBM energy in band gap corrected calculations, we obtain the charged defect formation energies of $D^q$ in $Mg_2(Si_{1-x}Sn_x)$ at $E_F = E_{CBM}$ by interpolating the defect formation energies of $D^q$ at $E_F=E_{CBM}$ in $Mg_2Si$ and $Mg_2Sn$. The position of the VBM is estimated using the corrected band gap obtained from the HSE calculations of 96-atom $Mg_2(Si,Sn)$ supercells.

**Band gap correction.** For defect formation energy calculations where the band gap is important, we used the hybrid-DFT without SOI. However, for the thermoelectric coefficient calculations, we used the hybrid-DFT band gap corrected PBE+SOI band structures. To obtain the band gaps of alloy compositions, we perform the 96-atom supercell calculations. Note that due to the conduction band convergence of $Mg_2Si_{1-x}Sn_x$ near x = ~0.5 to 0.6, the band gap is piecewise linear on x. We also test the virtual crystal approximation (VCA) [30] to predict the band gaps. But it is found that the band gaps from VCA are underestimated compared to the supercell approach.

**Phonon band structure.** We calculated the phonon band structure of $Mg_2Si$ and $Mg_2Sn$ using the supercell method implemented in the PHONOPY code [31,32]. The force constants of atoms were calculated using the atomic position perturbed 96-atom supercells within HSE06 or PBE calculations. At this moment, the atomic displacement in perturbed supercells was set to 0.01 Å.

**Defect density.** The defect density of defect $D^q$ [$n(D^q)$] in materials can be computed using the defect formation energy $E_{\text{FORM}}$ and the Boltzmann factor, considering the equilibrium conditions of defect generation temperature $T$ [26]. As the defect formation energies are a function of atomic chemical potential and the Fermi level during defect generation, the defect densities also depend on the atomic chemical potential, Fermi level, and $T$ given as

$$n(\mu_i, E_{Fermi}|D^q, \text{Mg}_2\text{X}) = n_{latt}\theta_{deg} \exp\left(-\frac{H_{FORM}(D^q)}{k_BT}\right) \simeq n_{latt}\theta_{deg} \exp\left(-\frac{E_{FORM}(D^q)}{k_BT}\right)$$

where $n_{\text{latt}}$ is the number density of available lattice sites in materials, $\theta_{\text{deg}}$ is the number of degrees of internal freedom of the defect on a lattice site, $H_{\text{FORM}}$ is the formation Enthalpy of the charged defect, and $k_B$ is the Boltzmann constant. Here we assume that the volume expansion effects for defects are negligible and thereby we use the defect formation energy instead of the formation Enthalpy. At last, the defect density of $D^q$ in Mg$_2$X are determined when atomic chemical potential and Fermi level are determined.

At a finite and non-zero temperature, the electrons in materials can be thermally activated. Thus, in gapped materials, the electron densities at the conduction bands ($n_C$) and the hole densities at the valence bands ($n_V$) are also determined when Fermi level and temperature are given as

$$(n_C - n_V)|_{E_{Fermi}} = \int_{E<\infty} g(E) \cdot f(E|E_{Fermi}) \, dE$$

where $g(E)$ is the density of states and $f(E)$ is the Fermi-Dirac distribution $f(E) = 1/(\exp((E - E_{\text{Fermi}})/k_BT) + 1)$. The g(E) is calculated using DFT, PAW, PBE level with spin-orbit interaction (SOI). Note that the inclusion of SOI plays a crucial role in thermoelectric transport coefficients due to its significant effect on band dispersions [33]. Here, to overcome the severe bipolar nature, we correct the band gap using the HSE06 band gap. The left hand side of the equation is called "Effective doping density". Since the whole system should be charge-neutral, the total of charge densities from free carriers and defect densities should be zero. From this charge neutrality condition, the charge-neutrality level (**CNL**) is determined from the Fermi level which satisfies the following

equation

$$n_C - n_V = \Sigma_{D,q}\, q \cdot n(D^q).$$

If the charge state q is given, the defect density is computed only for the given q. However, if the charge state q is not given for defect, the defect density means a sum of all defect densities of q = 2+, 1, 0, 1-, 2-.

**Doping efficiency.** We also define the doping efficiency (*e*) of $Mg_2X$ as the ratio of the effective doping density over the whole defect density as

$$e = \frac{|\Sigma_{D,q}\, q \cdot n(D^q)|}{2\, \Sigma_{D,q}\, n(D^q)}.$$

Thus, as the major defects in $Mg_2X$ are double donors or acceptors, the doping efficiency is normalized by 2. Finally, the doping efficiency of $Mg_2X$ would be between 0% and 100%.

**Semi-classical Boltzmann Transport calculation.** We compute the Boltzmann transport equation using the DFT-PBE band structures with the SOI effect. To solve the Boltzmann transport equation, we use the BoltzTraP code [34]. For the band dispersion description, we include the SOI in the band structure of alloy within the VCA approximation. For the band gap correction, we use the band gaps obtained from the hybrid-DFT band gaps obtained from the supercell approach without the SOI.

### 3. Results and Discussion

The hybrid-DFT calculations well reproduce the band structure, especially for the band gap. We find no big difference in band structure shapes between HSE06, PBE, and modified Becke-Johnson potential (mBG): see Figure S1 in Supporting Material (SM). We found that the low-energy band structures are very similar to each other. While the band gap of PBE is underestimated, the band gaps of HSE06 and mBJ well reproduce the experimental gaps. Also note that, in our calculations, the HSE06 and mBJ have very similar band gaps. For $Mg_2Si$, the band gaps are 0.200 eV, 0.571 eV, 0.593 eV from PBE, HSE06, and mBJ, respectively. The inclusion of the SOI *negligibly* affect the band gap of $Mg_2Si$: 0.190 eV, 0.559 eV, and 0.583 eV from PBE, HSE06, and mBJ, respectively. However, the SOI effect on band gap is *significant* for $Mg_2Sn$. Without SOI, the band gaps are -0.192 eV, 0.145 eV, and 0.150 eV, from PBE, HSE06, and MBJ, respectively. After inclusion of SOI, the band gaps are -0.349 eV, -0.0347 eV, and -0.002 eV from PBE, HSE06, and MBJ, respectively. However, the band gaps are negative or negligible when we use the SOI. The mBJ also well reproduce the band gaps. However, the mBJ does not give total energies. Since the band gap is critical for defect charge stability and we need total energy for formation energy calculations, we used the HSE06 band gaps without SOI.

Our hybrid-DFT calculations also well reproduce the stability of $Mg_2Si$ and $Mg_2Sn$, as compared with DFT calculations (see Table 1). The heats of formation energy are computed as -0.189 ev/atom and -0.267 eV/atom for $Mg_2Si$ and $Mg_2Sn$ respectively, which are comparable to the to the experimentally observed values (-0.224 and -0.276 eV/atom) [35]. However, in DFT-PBE calculations, the stability is underestimated: -0.166 eV/atom for $Mg_2Si$ and -0.192 eV/atom for $Mg_2Sn$.

Next we tested the lattice stability of $Mg_2Si$ and $Mg_2Sn$ by calculating the phonon band structure of them: see Figure S2 in SM. The HSE06 and PBE phonon band structures are very similar. We found no negative phonon mode, indicating that the lattice structures of $Mg_2Si$ and $Mg_2Sn$ are dynamically stable. But we found big difference in phonon frequencies of $Mg_2Si$ and $Mg_2Sn$. For $Mg_2Si$, the phonon modes range from 0 to ~11 THz without phonon band gap. However, for $Mg_2Sn$, the phonon frequencies are smaller than for $Mg_2Si$ due to the heavier atomic mass of Sn than that of Si, ranging from 0 to ~9 THz. And there is a small phonon band gap near 4 THz.

Figure 1 shows the charged defect formation energies of native point defects in Mg$_2$Si and Mg$_2$Sn for Mg-rich and Mg-poor conditions. In these materials, the most dominant defects are I$_{Mg}^{2+}$ donors and V$_{Mg}^{2-}$ acceptors. As their charge states are fixed when $E_{Fermi}$ changes in the range inside the band gap, they are shallow donors and acceptors, respectively. Compared to conventional DFT, the VBM positions are lowered in the hybrid-DFT calculations [20]. While the DFT calculations gives the defect formation energy of I$_{Mg}^{2+}$ at the VBM to be about 0.5 eV [18], our hybrid calculations show the formation energy of I$_{Mg}^{2+}$ at the VBM to be about 0 eV. Thus, in hybrid-DFT, the $E_g$ corrections lead to the enhanced stability of I$_{Mg}^{2+}$ for $p$-type conditions. For I$_{Mg}^{2+}$ and V$_{Mg}^{2-}$ in Mg$_2$Si, note that their $E_{FORM}$s are crossed at the $E_{Fermi}$ above the middle of the $E_g$, independent to the Mg chemical potential ($\mu_{Mg}$). These results of hybrid calculations indicate Mg$_2$Si is an unintentional $n$-type semiconductor in the binary phase. Note that, as the crossing point is located nearer the VBM in hybrid-DFT than in DFT, the hybrid-DFT predicts stronger n-type characteristics of Mg$_2$Si than DFT. Similarly, in Mg-rich Mg$_2$Sn the most stable defects are I$_{Mg}^{2+}$ donors, indicating $n$-type behavior in Mg-rich Mg$_2$Sn. However, in Mg-poor Mg$_2$Sn, the most stable defect configuration is the V$_{Mg}^{2-}$ acceptors, consistent to the experimentally reported $p$-type behavior in Mg$_2$Sn [10]. On the other hand, in Mg-poor Mg$_2$Sn, the $E_{FORM}$ of Sn$_{Mg}^{1+}$ is comparable to V$_{Mg}^{2-}$. Thus, although V$_{Mg}^{2-}$ generates holes in Mg-poor Mg$_2$Sn, many hole carriers can be compensated by this charge compensation center (Sn$_{Mg}^{1+}$).

We analyzed the geometries of two major defects I$_{Mg}$ and V$_{Mg}$ in Mg$_2$Si and found that the structural changes are only dominant for the nearest neighbor (n.n) of the defect and next n.n (n.n.n) atoms (see Table 2). For I$_{Mg}^{2+}$ in Mg$_2$Si, there are 8 n.n Mg atoms and 6 n.n.n Si atoms. The distance between I$_{Mg}$ and n.n Mg atoms are 2.750 Å before structure relaxation. After structural relaxation, the n.n Mg atoms move outward direction by 0.163 Å. Meanwhile, the distance between I$_{Mg}$ and n.n.n Si atoms is chaned by -0.087 Å from 3.175 Å. Note that the atomic movement near the defect can be understood by the electrostatic direction. And the structural relaxation is larger for Mg-neighbor atoms compared to Si-neighbor atoms. For $V_{Mg}^{2-}$, the distances to four n.n Si and six n.n.n Mg atoms are changed by 0.113 and -0.202 from 2.750 and 3.175 Å, respectively. The structural change is slightly enhanced in the charged configurations compared to neutral states. The structural change in HSE06 is slightly larger than the change in PBE.

We also analyzed the defect geometries of $I_{Mg}$ and $V_{Mg}$ in Mg2Sn (see Table 3). The structural relaxations are also occurred. When the neighboring atoms have opposite sign of charge compared to the defect, the neighboring atoms move outward directions, while atoms having same atomic charges move outward direction. Also note that, there are significant differences between defect configurations from HSE06 and PBE.

We predicted the defect formation energies of $Mg_2(Si_{1-x}Sn_x)$ solid solutions (x = 0 to 1.0) [see Supporting Material (SM)]. Considering the nearly preserved position of the CBM states after $E_g$ correction by hybrid-functional calculations, we estimated the $E_{FORM}[d^q]$ in $Mg_2(Si_{1-x}Sn_x)$ at $E_{Fermi}$ = $E_{CBM}$ by interpolating the $E_{FORM}[d^q]$ at $E_{Fermi}$ = $E_{CBM}$ in Mg2Si and Mg2Sn. Meanwhile, the $E_g$s of solid solutions were predicted by performing hybrid-functional calculations within the virtual-crystal approximation [28]. In Mg-rich solid solutions, $I_{Mg}$ are the main source of $n$-type conductivity. Furthermore, the larger $E_g$ of solid solutions compared to Mg2Sn leads to the weakened $p$-type behavior due to the increased $E_{FORM}$ of $V_{Mg}$ acceptors near the VBM.

Figure 2 show the calculated defect densities of native point defects in Mg2(Si,Sn) using the charge neutrality conditions and the defect formation probability at a temperature of 800 K [26] (see SM). In Mg-rich Mg2Si, the defect density of $I_{Mg}$ was $3 \times 10^{18}$ cm$^{-3}$ whereas that of $V_{Mg}$ is only $8 \times 10^{15}$ cm$^{-3}$. Even in Mg-poor Mg2Si, the density of $V_{Mg}$ ($7 \times 10^{16}$ cm$^{-3}$) is still smaller than that of $I_{Mg}$ ($3.1 \times 10^{17}$ cm$^{-3}$), indicating the unintentional $n$-type conductivity in Mg2Si independent of $\mu_{Mg}$. Thus, by changing the $\mu_{Mg}$ from an Mg-rich to an Mg-poor condition, the effective doping density, which is defined as $n_C - n_V$, changes from $5.8 \times 10^{18}$ cm$^{-3}$ to $4.8 \times 10^{17}$ cm$^{-3}$, where $n_c$ and $n_v$ are the charge densities for free electrons and holes, respectively. However, in Mg$_2$Sn, due to the increased defect density of V$_{Mg}^{2-}$, the majority of charged defects turns from I$_{Mg}$ to V$_{Mg}$ going from an Mg-rich to an Mg-poor condition. In an Mg-poor condition the density of V$_{Mg}$ can reach $3.3 \times 10^{19}$ cm$^{-3}$ and the material can be *p*-type. At this moment the densities of positive charged Sn$_{Mg}^{1+}$ and Sn$_{Mg}^{2+}$ defects ($1.3 \times 10^{19}$ cm$^{-3}$ and $5.2 \times 10^{18}$ cm$^{-3}$) are comparable to that of V$_{Mg}^{2-}$ ($2.4 \times 10^{19}$ cm$^{-3}$), indicating the severe charge compensation between native donors and acceptors. As a result, the $n_C - n_V$ can be $-3.1 \times 10^{19}$ cm$^{-3}$. Similar to Mg$_2$Sn, *p*-type conduction can be achieved in Mg$_2$(Si$_{1-x}$Sn$_x$) solid solution (x ≥ 0.8) under Mg-poor conditions. However, the charge compensation effect exists for *p*-type due to the next most probable defect I$_{Mg}^{2+}$ in Mg-poor solid solutions.

By comparing the effective doping density $n_C - n_V$ with the defect densities in Mg$_2$(Si,Sn), we quantified the doping efficiency of native defects for *p*-type Mg$_2$(Si,Sn) (Figure 3). Here, the defect doping efficiency is defined as the ratio of the effective doping density to the sum of all defect densities with all charge states for the major defect. In Mg$_2$Si, the large defect formation energy and small defect density of acceptors are responsible for the unintentional *n*-type conductivity with the effective doping density being less than $\sim 6 \times 10^{19}$ cm$^{-3}$. The *n*-type doping efficiency of I$_{Mg}$ in Mg$_2$Si is 97% and 61% for Mg-rich and Mg-poor conditions, respectively. In Mg$_2$Sn, the *p*-type conduction is possible with Mg-poor conditions. The electron and hole densities can exceed $10^{19}$ cm$^{-3}$ with the formation of I$_{Mg}$ and V$_{Mg}$, respectively, depending on $\mu_{Mg}$. However, due to the charge compensation effect, *p*-type doping efficiency of V$_{Mg}$ is significantly reduced. For Mg-poor Mg$_2$Sn, *p*-type doping efficiency of V$_{Mg}$ is only 24% with an effective doping density of $-3.0 \times 10^{19}$ cm$^{-3}$. With optimal $\mu_{Mg}$, *p*-type doping efficiency of V$_{Mg}$ can be increased to ~60%. We found that *p*-type doping efficiency by native defects can be enhanced in Mg$_2$(Si$_{1-x}$Sn$_x$) solid solution systems. With increased $E_g$ and increased counter defect formation energy, the charge compensation effect can be weakened. As shown in Figure 3(b), in the solid solution composition of $Mg_2Si_{0.1}Sn_{0.9}$, the highest *p*-type doping efficiency of 48% can be achieved with an effective doping density of -1.0 × 10$^{19}$ cm$^{-3}$.

From the semi-classical Boltzmann transport theory combined with the first-principles band structures of $Mg_2(Si,Sn)$, we estimated the optimal doping concentration for thermoelectric power factors of $Mg_2Si$, $Mg_2Sn$, and their solid solutions (Figure 4 and SM). We used the constant-relaxation time approximation given as $\tau = \left(\frac{300K}{T}\right) \times 10^{-14} s$ [36]. For $Mg_2(Si,Sn)$, the optimal effective doping density is $10^{20} - 10^{21}$ cm$^{-3}$, consistent with other reports [37,38], which is much larger than the possible *p*-type hole densities and *n*-type electron densities [Figure 3(a)]. Our results indicate that, although the alloying of $Mg_2Si$ with $Mg_2Sn$ can make it a *p*-type semiconductor with relatively high doping efficiency, it still needs an extrinsic dopant to increase the carrier concentration to optimize the power factor. It is also found that the *p*-type power factor can be higher in $Mg_2Si$ than in $Mg_2Sn$ due to the smaller valence band effective mass in $Mg_2Sn$ by the spin-orbit interaction induced valence band splitting.

Additionally, we examined thermal stability of *p*-type conductivity for $Mg_2Si$-based semiconductors. We performed the nudged elastic band calculations for the diffusion of Mg interstitial defects in $Mg_2Si$ [Figure 4(b)] [29]. The migration barriers ($E_{mig}$) of $I_{Mg}$ is 801 meV when $I_{Mg}$ migrates via an *interstitialcy* diffusion mechanism. The diffusion coefficient of $I_{Mg}$ is calculated using the $E_{FORM}$, $E_{mig}$, and the vibrational frequencies of $I_{Mg}$ [29,39]. In Mg-rich *n*-type $Mg_2Si$, the *D* is 3.7 × 10$^{-19}$ m$^2$/s at 800 K when the $E_{Fermi}$ was at the CBM. From the *D* and diffusion time *t*, we estimated the diffusion length *L* as $L = \sqrt{6Dt}$ [27]. For 1000 hours under the annealing temperature of 800 K, $I_{Mg}$ can diffuse to 2.8 μm, which is comparable to the grain size of polycrystalline $Mg_2Si$ [33,40]. In this Mg-poor condition, the *D* can be highly enhanced to 1.9 × 10$^{-15}$ m$^2$/s at 800 K with the lowered defect formation energy. The *L* of 6.4 μm is possible for 1 hour under at 800 K.

Our results imply a thermal instability of the thermoelectric properties due to formation of native defects at thermoelectric working temperature. Although we can make *p*-type solid solutions by controlling the $\mu_{Mg}$ or by doping suitable *p*-type elements, the acceptors can be self-compensated by the generation and diffusion of $I_{Mg}$ defects from outside. Thus, for stable *p*-type conduction, very low acceptor formation energies are required to overcome the defect compensation by native donor states ($I_{Mg}^{2+}$ and $Sn_{Mg}^{1+}$).

## 4. Conclusions

In summary, we investigated the native point defects in $Mg_2Si$, $Mg_2Sn$, and their solid solutions. We adopted the hybrid-functional to overcome the band gap problem in DFT. There are abundant intrinsic defects in $Mg_2Si$ such as Mg interstitials and Mg vacancies. While $Mg_2Si$ is unintentional *n*-type, the $Mg_2(Si,Sn)$ can be tuned from *n*-type to *p*-type going from Mg-rich to Mg-poor conditions. As $Mg_2Sn$ has a low *p*-type doping efficiency due to the severe charge compensation effect between donors and acceptors, an alloying strategy between $Mg_2Si$ and $Mg_2Sn$ might be required to obtain a higher *p*-type doping efficiency. However, the possible *p*-type carrier density is still low compared to the optimal carrier concentration for thermoelectric power factor. Moreover, the fast diffusion of Mg interstitial can degrade the thermal stability of *p*-type samples. Based on the results, we suggest that an extrinsic impurity doping strategy is required for $Mg_2(Si,Sn)$ thermoelectric materials to stabilize the hole high doping concentration with high doping efficiency. Further hybrid-density functional studies on impurity doped $Mg_2(Si,Sn)$ will help to find suitable *p*-type impurity dopants.

## ACKNOWLEDGEMENT

This work was supported by the International Energy Joint R&D Program of the Korea Institute of Energy Technology Evaluation and Planning (KETEP) granted from the Ministry of Trade, Industry & Energy (MOTIE), Republic of Korea: Nos. 20188550000290, 20172010000830). It was also supported by the Korea Electrotechnology Research Institute (KERI) Primary Research Program through the National Research Council of Science and Technology (NST) funded by the Ministry of Science and ICT (MSIT) of the Republic of Korea (No. 20A01025). Also, one of the authors (JdB) was partially funded by the Deutsche Forschungsgemeinschaft (DFG, German Research Foundation) - project number 396709363.

# Tables

**Table 1.** Calculated formation enthalpies for $Mg_2Si$ and $Mg_2Sn$.

| Phase | Formation enthalpy (eV/atom) | | |
|---|---|---|---|
| | Hybrid-DFT, HSE06 | DFT, GGA PBE | Experiment |
| $Mg_2Si$ | -0.189 | -0.166 | -0.224[a] |
| $Mg_2Sn$ | -0.267 | -0.192 | -0.276[a] |

[a] Reference [34]

**Table 2.** Defect geometries of $I_{Mg}$ and $V_{Mg}$ in $Mg_2Si$. The distance from a defect to a neighboring atom is summarized.

| Defect in $Mg_2Si$ | Defect charge | Distance from defect | | Distance change after relaxation | | |
|---|---|---|---|---|---|---|
| | | Neighbor | Before relaxation | PBE | HSE06 | |
| $I_{Mg}$ | 2+ | 8 n.n Mg atoms | 2.750 | +0.162 | 0.163 | Outward |
| | | 6 n.n.n Si atoms | 3.175 | -0.084 | -0.087 | Inward |
| | 0 | 8 n.n Mg atoms | 2.750 | 0.147 | 0.150 | Outward |
| | | 6 n.n.n Si atoms | 3.175 | -0.064 | -0.070 | Inward |
| $V_{Mg}$ | 2+ | 4 n.n Si atoms | 2.750 | 0.106 | 0.113 | Outward |
| | | 6 n.n.n Mg atoms | 3.175 | -0.194 | -0.202 | Inward |
| | 0 | 4 n.n Si atoms | 2.750 | 0.102 | 0.105 | Outward |
| | | 6 n.n.n Mg atoms | 3.175 | -0.186 | -0.191 | Inward |

**Table 3.** Defect geometries of $I_{Mg}$ and $V_{Mg}$ in $Mg_2Sn$. The distance from a defect to a neighboring atom is summarized.

| Defect in $Mg_2Sn$ | Defect charge | Distance from defect | | Distance change after relaxation | | |
|---|---|---|---|---|---|---|
| | | Neighbor | Before relaxation | PBE | HSE06 | |
| $I_{Mg}$ | 2+ | 8 n.n Mg atoms | 2.923 | +0.153 | +0.153 | Outward |
| | | 6 n.n.n Si atoms | 3.375 | -0.065 | -0.071 | Inward |
| | 0 | 8 n.n Mg atoms | 2.923 | +0.127 | +0.142 | Outward |
| | | 6 n.n.n Si atoms | 3.375 | -0.043 | -0.063 | Inward |
| $V_{Mg}$ | 2+ | 4 n.n Si atoms | 2.923 | +0.020 | +0.049 | Outward |
| | | 6 n.n.n Mg atoms | 3.375 | -0.122 | -0.146 | Inward |
| | 0 | 4 n.n Si atoms | 2.923 | 0.018 | 0.029 | Outward |
| | | 6 n.n.n Mg atoms | 3.375 | -0.103 | -0.128 | Inward |

**Figure Captions**

Fig. 1 $E_{\text{FORM}}$ curves for native point defects in (a) $Mg_2Si$ under Mg-rich, (b) $Mg_2Si$ under Mg-poor, (c) $Mg_2Sn$ under Mg-rich, and (d) $Mg_2Sn$ under Mg-poor conditions.

Fig. 2 Densities of native point defects in (a) $Mg_2Si$, (b) $Mg_2(Si_{0.3}Sn_{0.7})$, (c) $Mg_2(Si_{0.2}Sn_{0.8})$, and (d) $Mg_2Sn$ are drawn as a function of Mg chemical potential from Mg-poor to Mg-rich.

Fig. 3 Contour maps of (a) the calculated effective doping density and (b) doping efficiency of $Mg_2(Si,Sn)$ are drawn as a function of the composition and the Mg chemical potential.

Fig. 4 Calculated power factors for (a) $Mg_2Si$ and (b) $Mg_2Sn$ are drawn as a function of carrier concentration for various temperatures from 300 to 1000 K. The *p*-type power factor is shown as positive the *n*-type as negative.

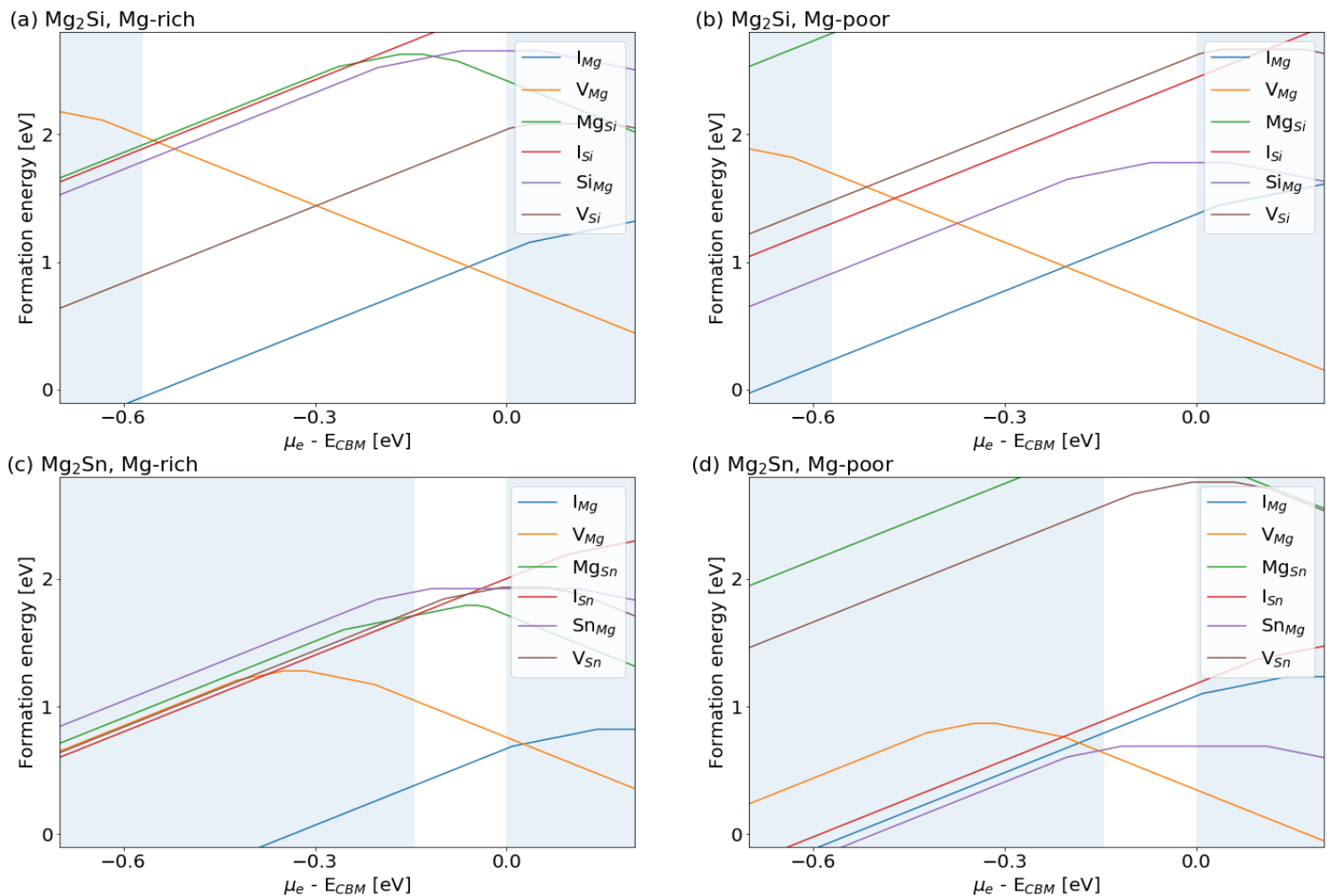

Figure 1

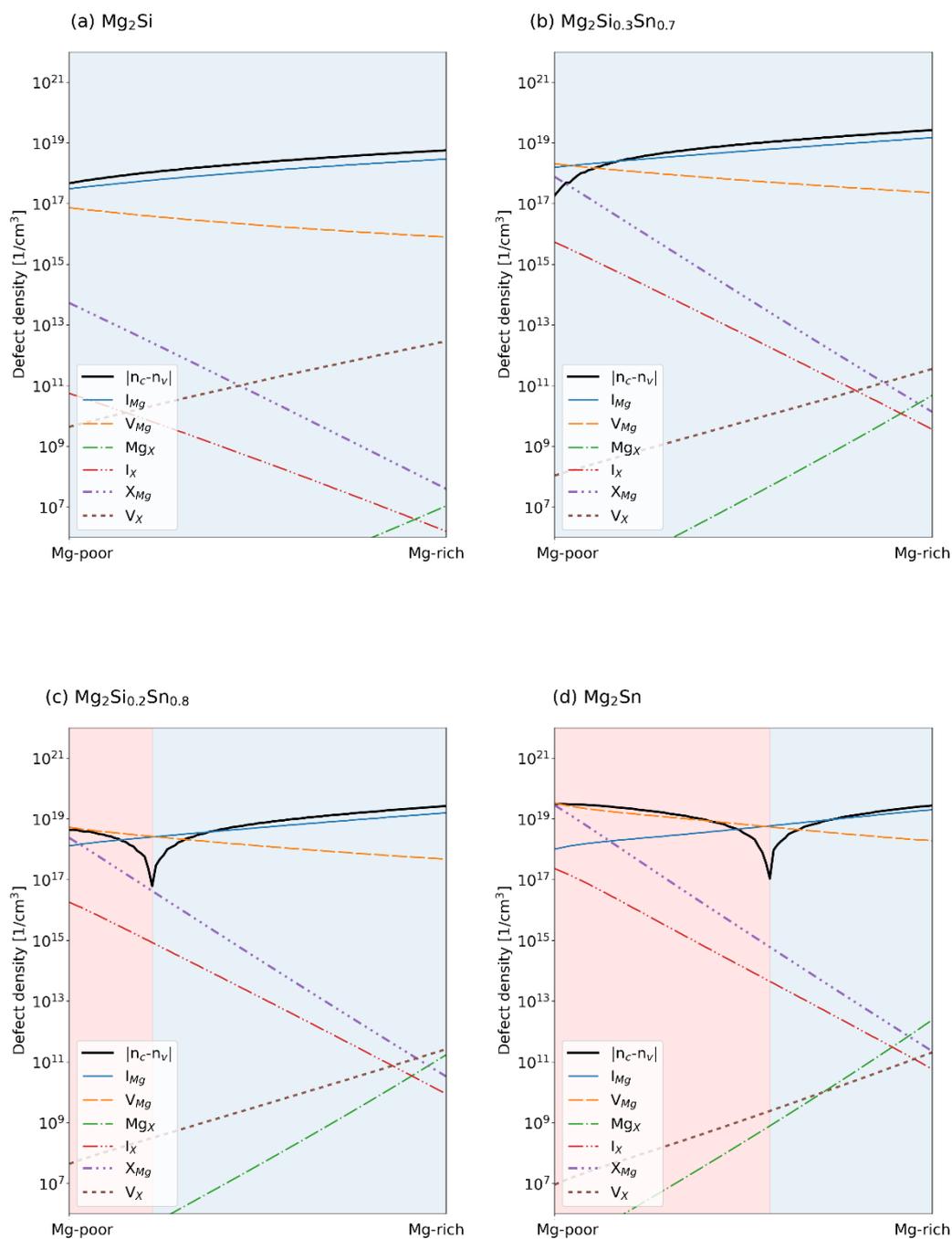

**Figure 2**

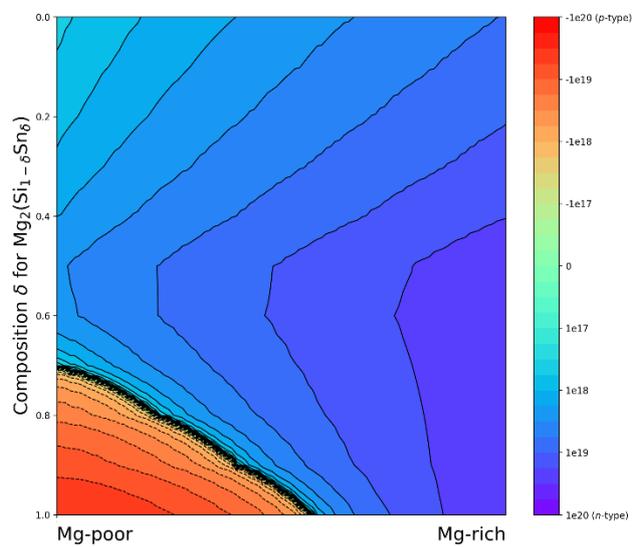
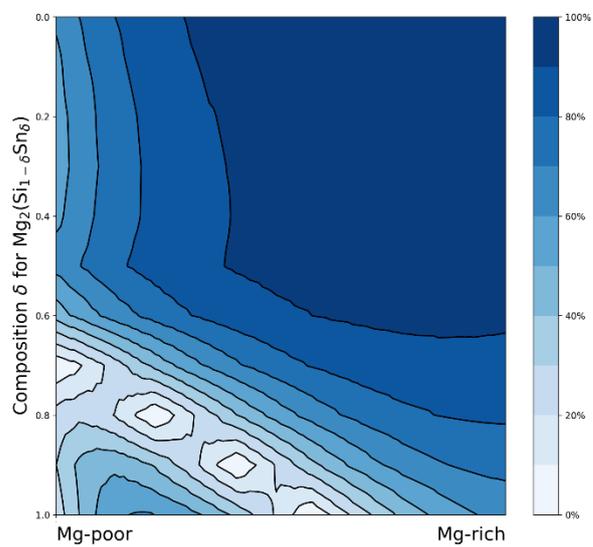

**Figure 3**

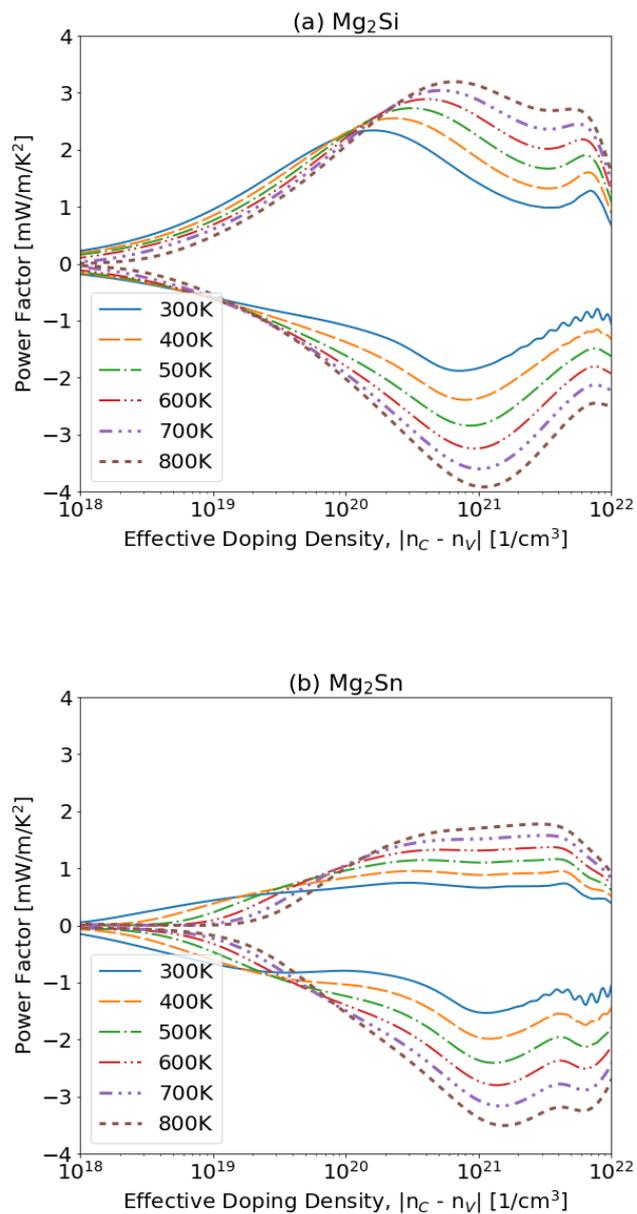

Figure 4